\pdfoutput=1
\documentclass[preprint,11pt]{elsarticle}

\usepackage{graphicx}
\usepackage{hyperref}           %
\usepackage[english]{babel}
\usepackage{lineno}

\usepackage[T1]{fontenc}       %
\usepackage[utf8]{inputenc} %

\newcommand{\AAA}{\mbox{\AA}}

\journal{Computer Physics Communications}

\graphicspath{{./}{figures/}}

\begin{document}

\begin{frontmatter}

\title{METAGUI 3: a graphical user interface for
  choosing the collective variables in molecular dynamics simulations}

\author{Toni Giorgino}
\ead{toni.giorgino@cnr.it}
\address{Institute of Neurosciences, National Research Council (CNR-IN),\\ Corso Stati Uniti 4, I-35127, Padua, Italy}

\author{Alessandro Laio}
\ead{laio@sissa.it}
\address{International School for Advanced Studies (SISSA),\\ Via Bonomea 265, I-34136 Trieste, Italy}

\author{Alex Rodriguez}
\ead{alexrod@sissa.it}
\address{International School for Advanced Studies (SISSA),\\ Via Bonomea 265, I-34136 Trieste, Italy}


\begin{abstract}

  Molecular dynamics (MD) simulations allow the exploration of the
  phase space of biopolymers through the integration of
  equations of motion of their constituent
  atoms. The analysis of  MD trajectories often relies
  on the choice of collective variables (CVs) along which the 
  dynamics of the system is projected. We developed a graphical
  user interface (GUI) for facilitating the interactive choice of 
  the appropriate CVs.
  The GUI allows: defining interactively new CVs;
  partitioning the configurations  into
  microstates characterized by similar values of the CVs; calculating the free energies of the microstates for both
  unbiased and biased (metadynamics) simulations; clustering the microstates in
  kinetic basins; visualizing the free energy
  landscape as a function of a subset of the CVs used for the analysis.
  A simple mouse click allows one to quickly inspect
  structures corresponding to specific points in the landscape.

\end{abstract}

\end{frontmatter}
\newpage \noindent
{\bf PROGRAM SUMMARY}

\begin{small}
\noindent
{\em Program Title:} METAGUI 3                                         \\
{\em Licensing provisions:} GPLv3                                    \\
{\em Programming language:} Tcl/Tk, Fortran                                   \\
{\em Journal reference of previous version: } METAGUI [1]                  \\
{\em Does the new version supersede the previous version?:} No   \\
{\em Nature of problem:} Choose the appropriate collective variables for
describing the thermodynamics and kinetics of
a biomolecular system through biased and unbiased molecular dynamics.\\
{\em Solution method:} Provide an environment to compute and visualize
free energy surfaces as a function of collective variables,
interactively defined.  \\
{\em Additional comments:} 
METAGUI 3 is not a standalone program but a plugin that provides
analysis features within VMD (version 1.9.2 or higher).
\\

\end{small}

\section{Introduction}

Molecular dynamics (MD) is a computationally intensive simulation technique
which is used to gather  insights into the structural,
thermodynamic and kinetic properties of 
systems modeled at the atomistic level. MD simulations have been extensively used to study, e.g.,
biologically-relevant conformational changes of proteins, their
interactions with small molecules, and with other proteins.  The
problem of achieving sufficient sampling of the phase space has been
tackled  by careful software parallelization~\cite{phillips_scalable_2005,pronk_gromacs_2013} and
accelerated processors~\cite{harvey_acemd:_2009}.
Moreover, the sampling problem can be addressed by
enhanced sampling
techniques, such as umbrella sampling~\cite{torrie_nonphysical_1977},
metadynamics~\cite{laio_metadynamics:_2008}, steered MD~\cite{giorgino_high-throughput_2011}, accelerated MD~\cite{hamelberg_accelerated_2004}, etc.
These approaches  
alter the Hamiltonian or the temperature of the system,  therefore  modifying
its kinetic properties. However, they allow the  reconstruction of  the free
energy landscape.

Most of the biased sampling methodologies require the modeler to
choose a few ``important'' 
collective variables (CV), along which the important dynamics of the
system is assumed to happen. CVs are arbitrary functions of the internal coordinates of the
system; software such as PLUMED~\cite{tribello_plumed_2014} and
Colvars~\cite{fiorin_using_2013} allows their specification them \emph{via}
concise symbolic expressions.
The choice of CVs is normally done before
performing the simulation~\cite{laio_escaping_2002}. However it is common practice to choose the CVs
for a production run  by
 performing an analysis on a preliminary simulation. 
A careful choice of CVs
is also needed in unbiased molecular dynamics simulations,  in order to
obtain insight on the relevant processes that possibly occurred during the
time evolution.

In general,  choosing the CVs that better describe a conformational change, even if this change can be observed by visualizing the MD trajectory, can be far from trivial. 
For these reasons, there is a need
for software interfaces that simplify their iterative definition and
evaluation~\cite{giorgino_plumed-gui:_2014,biarnes_metagui._2012}.

\section{Features}

We here present METAGUI 3, a plug-in providing a graphical user interface
(GUI) to construct thermodynamic and kinetic models of processes
simulated by large-scale MD. The GUI is based on  the well-known MD visualization
software VMD~\cite{humphrey_vmd:_1996}; it extends the
features of an earlier version~\cite{biarnes_metagui._2012}, exploiting in particular: (a) its ability to toggle between
two representations, structural and free energy, enabling the quick
inspection of the configurations associated to  specific values of the CVs; (b) a procedure for grouping together similar structures in {\em microstates}
based  only on the collective variables; and (c) a simplified and
guided workflow for the analysis of metadynamics results.  METAGUI 3
was extensively rewritten to provide the following new features,
which will be presented in detail in the next sections:

\begin{enumerate}
\item an interface for defining and adding interactively new collective variables;
\item new algorithms for finding the  microstates, namely $k$-medoids~\cite{kaufman_book_1990} and Daura's algorithm~\cite{daura_peptide_1999};
\item a reorganized graphical interface;
\item support for the analysis of unbiased MD trajectories;
\item a new clustering method~\cite{rodriguez_clustering_2014} for identifying kinetic basins.
\end{enumerate}

\begin{figure}
  \centering
  \includegraphics[width=.7\textwidth]{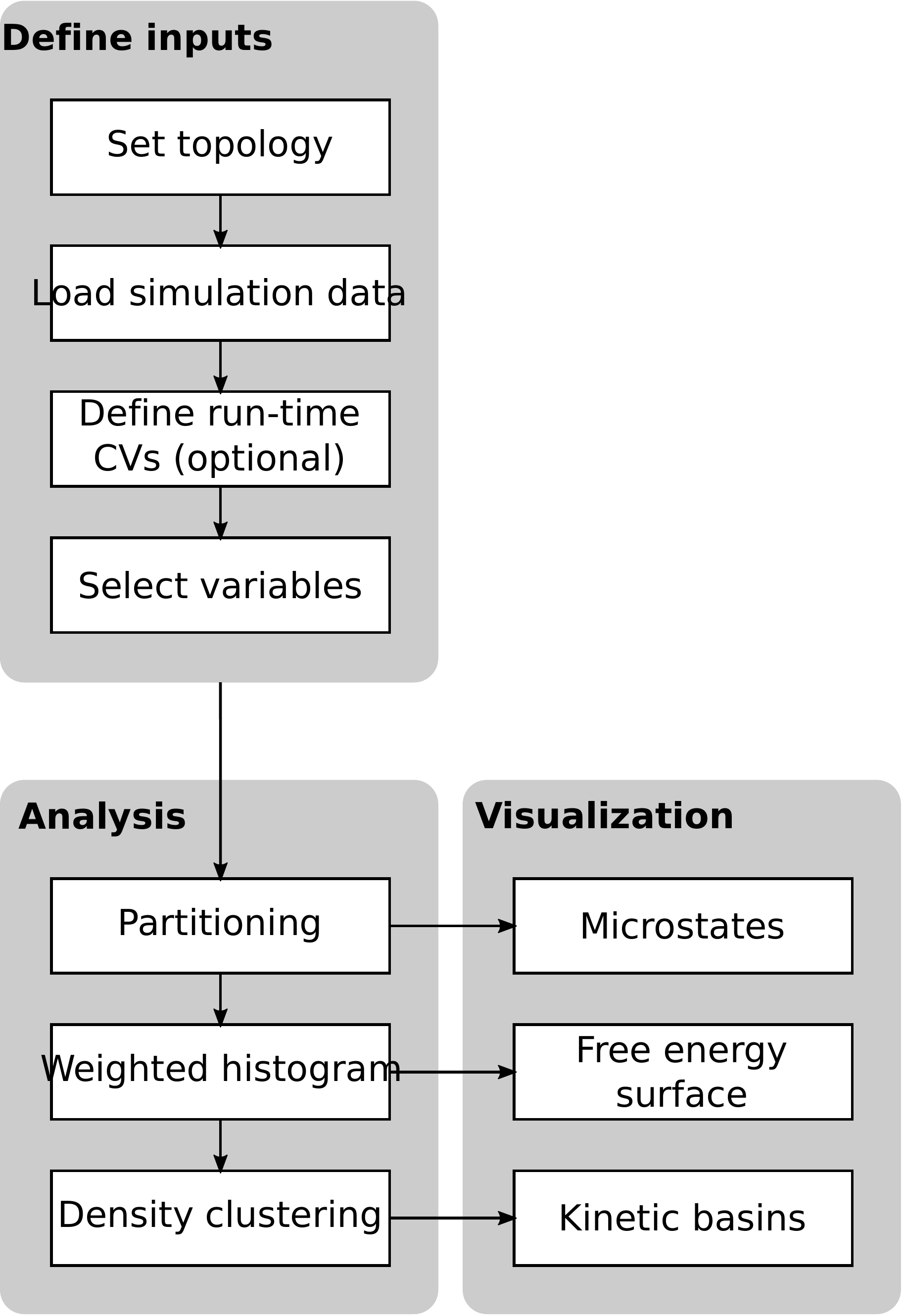}
  \caption{Overview of the analysis steps in METAGUI 3. The shaded
    boxes correspond to the three main sections of the GUI. The microstates,
    the free energies landscape, and the clusters can be
    visualized independently after the corresponding step in the
    analysis. }
  \label{fig:overview}
\end{figure}

\subsection{The graphical user interface}

Upon starting the tool, the user is presented with the METAGUI 3
window; the three main tasks are shown as tabs, namely \emph{Define
  inputs}, \emph{Analyze} and \emph{Visualize}. The three sections
correspond to the three logical steps of the analysis workflow shown
in Figure~\ref{fig:overview}.

\begin{figure}
  \centering
  \includegraphics[width=.9\textwidth]{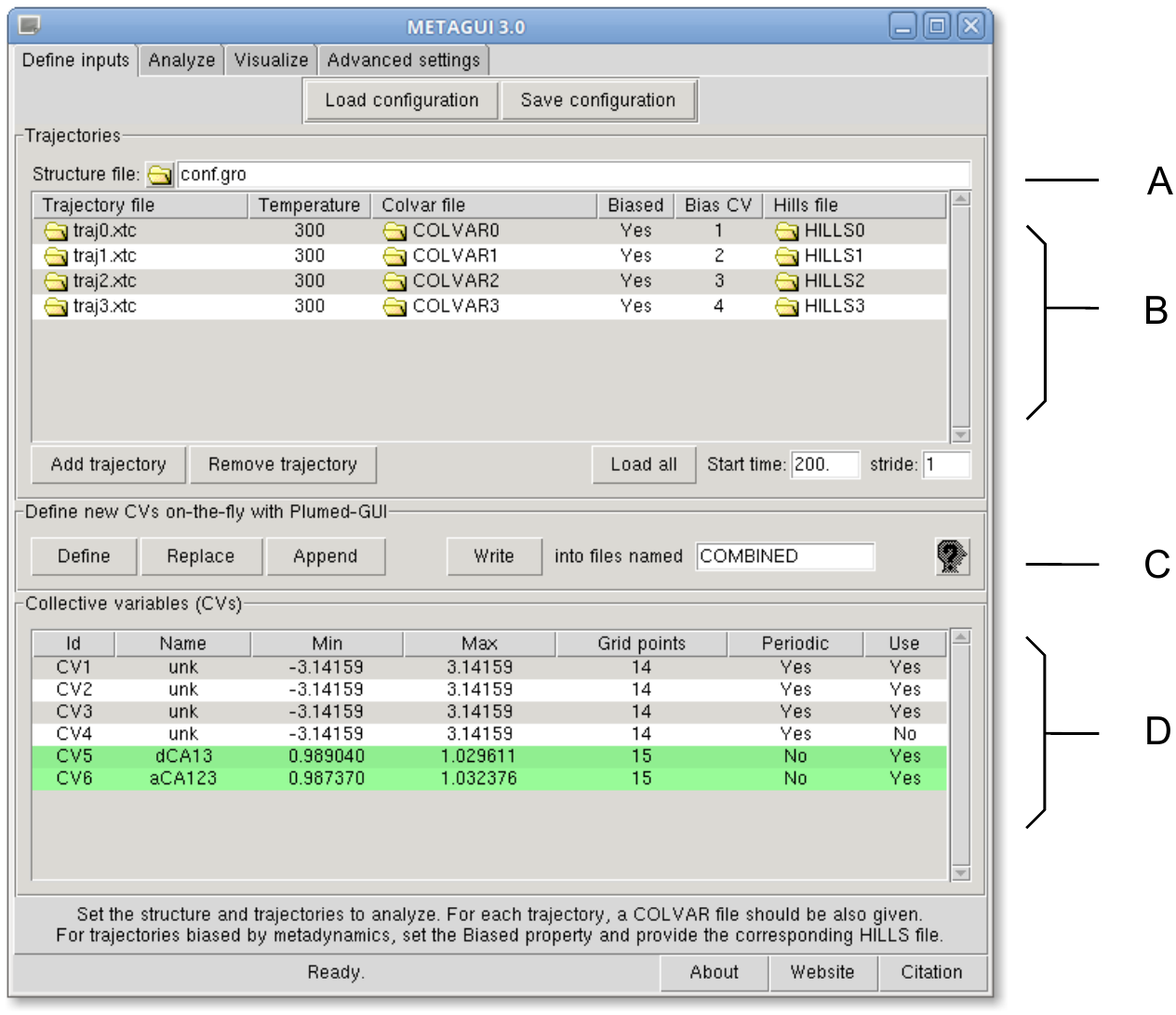}
  \caption{The \emph{Define inputs} tab, which allows setting the trajectory files, the system's topology (A) and  the temperatures of the available trajectories (B). The middle section (C) opens the Plumed-GUI collective-variable editor, allowing the run-time definition of unbiased CVs in addition to those originally  present in the \emph{COLVAR} files. Defined collective variables appear in section (D), either over a white background (CVs in the simulation files) or over a green 
background (CVs defined at run-time).}
  \label{fig:inputs}
\end{figure}

The \emph{Define inputs} tab (Figure~\ref{fig:inputs}) allows to
specify the input files, including the topology file of the system
(i.e.\ a file in PDB, PSF, or GRO format), and one or more trajectory files.
For each trajectory, a file containing the values of collective variables
(\emph{COLVAR} files) is required as well as the temperature of the
run; if the simulations are biased by metadynamics, the time-varying
potentials have to be supplied in matching \emph{HILLS}
files, produced by PLUMED-patched MD engines during the simulation.
The part of the plug-in dedicated to the analysis of biased simulations 
is unchanged with respect to the earlier version of METAGUI~\cite{biarnes_metagui._2012}. 
The \emph{Save} and \emph{Load configuration} buttons allow storing
and retrieving the whole set-up at once. 

Activating the \emph{Load all} button loads all the simulation data into
memory; this includes the molecular topology, all of the trajectory
frames, and the values of the collective variables. A summary of the
CVs appears on the lower-hand panel of the GUI, along with any
additional CV defined manually at runtime, as discussed in
Section~\ref{sec:run-time-definition}.

\begin{figure}
  \centering
  \includegraphics[width=1.0\textwidth]{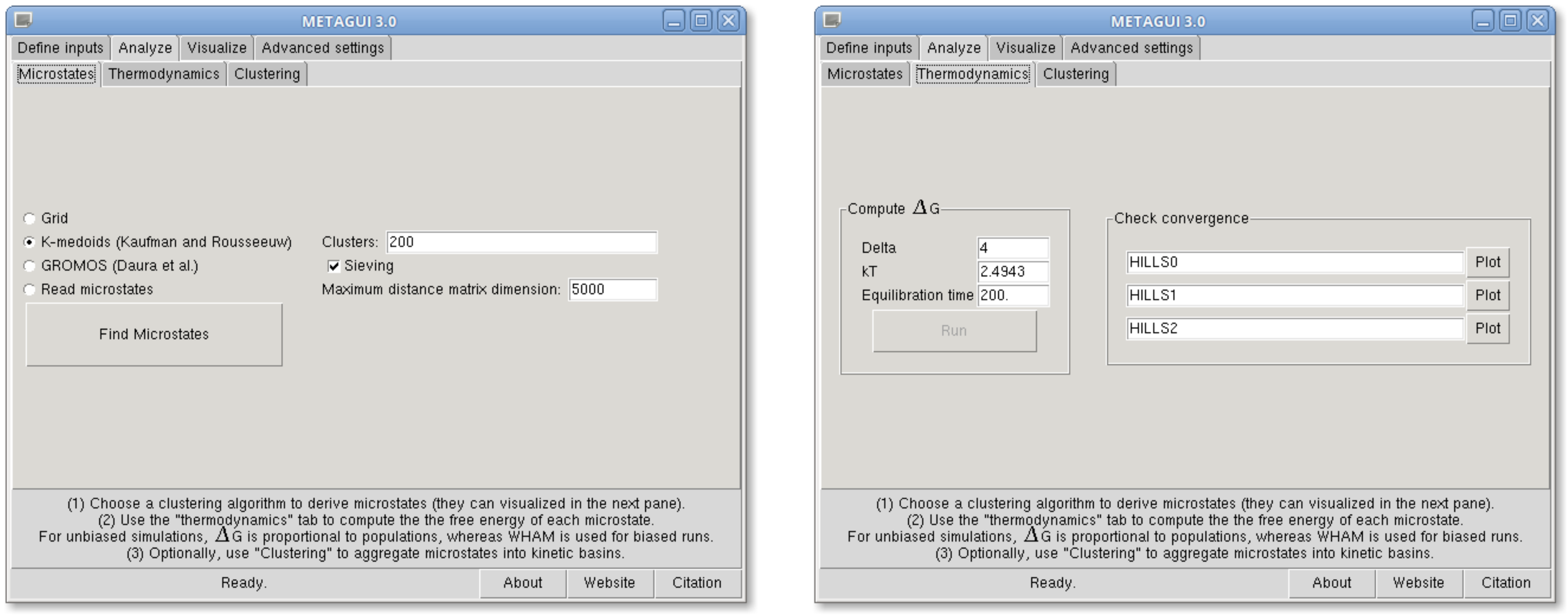}
  \caption{Two steps of the \emph{Analysis} procedure: (left) compute
    microstates via one of the available partitioning procedures; 
    (right) compute the thermodynamics of each microstate via either
    WHAM (biased trajectories) or Boltzmann
    inversion of the populations (unbiased).}
\label{fig:analyze}
\end{figure}

After the  input is defined, operations in the \emph{Analyze} tabs can
take place (Figure~\ref{fig:analyze}). In particular, the first step is
to group the configurations explored  during the simulations into
\emph{microstates}, namely sets of similar structures. Three partitioning algorithms are available, namely
(a) $k$-medoids with $k$-means++ initial seeding~\cite{kaufman_book_1990,vassilvitskii2006k} with or without
sieving~\cite{shao_clustering_2007}; (b) an implementation of the
algorithm by Daura et al.~\cite{daura_peptide_1999}; and (c) the
simple grid partitioning in CV space implemented in the earlier versions of
METAGUI.
The $k$-medoids correspond to a partitioning of
the space in $k$ Voronoi cells. The parameter $k$ directly defines
the number of microstates that are obtained.  Daura's
clustering algorithm~\cite{daura_peptide_1999} automatically determines the number 
of microstates by finding groups within a cut-off distance, that should be specified by the user.  
With respect to the grid partitioning, included also in this version, the use of more 
elaborated partition methods has the advantage of automatically focusing on 
the populated regions of the CV space. This avoids the exponential 
growth with the number of CVs of the  microstates with negligible population.

\begin{figure}
  \centering
  \includegraphics[width=.9\textwidth]{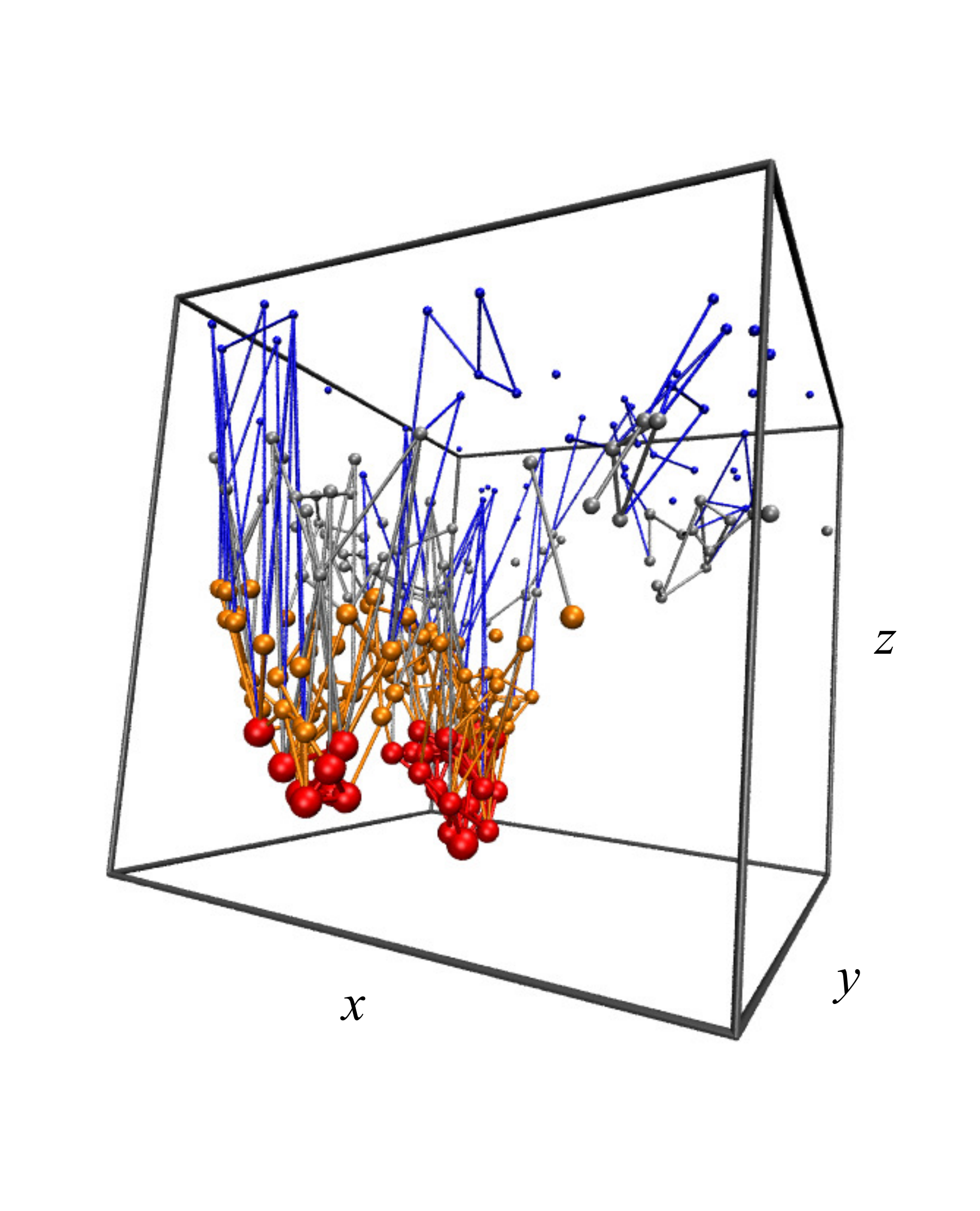}
  \caption{
    Free energy surface (FES) plotted with respect to two CVs
    ($x$ and $y$ axis). Each microstate is shown as a dot, whose color
    and $z$ coordinate reflect the corresponding $\Delta G$. Lines
    join kinetically-communicating microstates.}
\label{fig:fes}
\end{figure}

Once microstates have been computed, the tool allows their visualization.
The \emph{Visualize} tab lists the microstates along with their free
energy (Figure~\ref{fig:visualize-microstates}).  Selecting one of them it is possible to visualize the corresponding structures or
save them into a PDB file. It is also possible to seamlessly switch between a
collective variables space representation of microstates
and their atomic structure representation (Figure~\ref{fig:fes});
this greatly helps associating  structural rearrangements
with collective variable changes.

\subsection{Support for unbiased simulations}

A new feature in METAGUI 3.0 enables the analysis of trajectories resulting from
both unbiased and metadynamics-based simulations. It is possible to
switch between the two modes by toggling a \emph{Biased} check-box on
each trajectory. 
In  both biased and unbiased simulations the structures are grouped together
into a set of microstates, namely sets of structures with similar values of the
collective variables. For biased simulations,  free energies 
of the microstates  are then
computed by the
weighted histogram analysis method (WHAM)~\cite{roux_calculation_1995,Marinelli_2009}. For unbiased simulations,  free energies are
computed applying the Boltzmann inversion to the populations of the
microstates (visit counts).
The tool also allows analyzing simulations for which part of the trajectories are biased and part of the trajectories are not.

\subsection{Run-time definition of collective variables}\label{sec:run-time-definition}

\begin{figure}
  \centering
  \includegraphics[width=.9\textwidth]{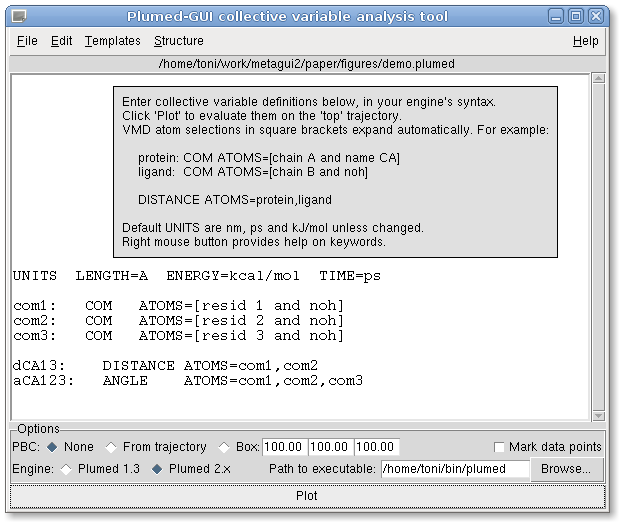}
  \caption{New collective variables can be defined via the Plumed-GUI
    interface \cite{giorgino_plumed-gui:_2014} and  included in the analysis at run-time. In
    this example, the user defined two runtime CVs in PLUMED 2.0
    syntax: the distance between the center of mass (\texttt{COM}) of
    the heavy atoms of residues 1 and 2; and the angle formed by the
    COM of residues 1, 2 and 3.}
\label{fig:plumed-gui}
\end{figure}

In METAGUI 3, PLUMED output can be read after the
simulation, in the form of \emph{COLVAR} files  as described
above, as well as recomputed at \emph{run-time}. This is an important improvement with respect to the previous version of the interface:  the user can now define
collective variables in the PLUMED language, and have them evaluated on
the loaded trajectories. CVs defined at run-time can either replace or be added to the
list of pre-computed CVs.  Definition of CVs at run-time is done
through the Plumed-GUI software (Figure~\ref{fig:plumed-gui}), which
provides a guided environment to prototype CV declarations with
templates and symbolic atom selection keywords~\cite{giorgino_plumed-gui:_2014}.
Plumed-GUI is distributed with VMD since version 1.9.2, and can be updated separately
from it.

The new version of METAGUI described in this work  can process the output of the PLUMED 1.3~\cite{bonomi_plumed:_2009} and 2.x~\cite{tribello_plumed_2014}
engines, making it compatible with a number of different molecular
dynamics packages like AMBER~\cite{case_amber_2005}, NAMD~\cite{phillips_scalable_2005},
GROMACS~\cite{pronk_gromacs_2013} and several others.
The use of CVs computed at run-time only requires
PLUMED's \emph{driver}, a stand-alone
executable which is built independently of MD engines.

\subsection{Clustering strategies}

METAGUI 3 allows quickly detecting the presence of metastable states by applying the 
Density Peak clustering approach~\cite{rodriguez_clustering_2014}. The approach is based 
on the assumption that cluster centers are surrounded by neighbors with lower local 
density and that they are at a relatively large distance from any points with a higher 
local density. The method proceeds by computing two quantities for each
point: the density $\rho_i$ and the minimum distance to a point with higher density 
$\delta_i$. In its simplest formulation, the density is estimated by the number of 
neighbors within a cut-off, although more elaborated definitions can be employed. 
Once computed, both quantities are plotted in the so-called decision plane. 
Cluster centers stand out naturally as outliers on the graph and can be manually picked up.
After the cluster centers have been found, each remaining point is assigned to the 
same cluster as its nearest neighbor of higher density.

In the context of molecular dynamics, the quantity $\rho$ has a direct physical interpretation
through the Boltzmann equation, which relates the logarithm of the probability density with the
free energy. Therefore, in the version implemented in METAGUI 3, the density of a microstate
is computed by $\rho_i=\exp\left(-\Delta G_i / k_BT \right)$, where $k_B$ is the Boltzmann constant and $T$ is the temperature. This definition leads to realistic
density values even in the case of biased simulations, where standard density calculation 
methods cannot be employed. The distance between microstates is computed as the Euclidean distance 
in the CVs space. However, to deal with the different ranges of variation of the CVs, it is necessary to
include some kind of normalization. This is done by dividing the coordinates by the bin size in the grid. 
Therefore, the distance can be interpreted as the number of hypercubes between microstates. 
The clusters obtained by this procedure are a good approximation   of the metastable
states normally obtained by Markov State Modeling. Therefore, they can be used 
as a starting point for kinetic analysis.

\begin{figure}
  \centering
  \includegraphics[width=.9\textwidth]{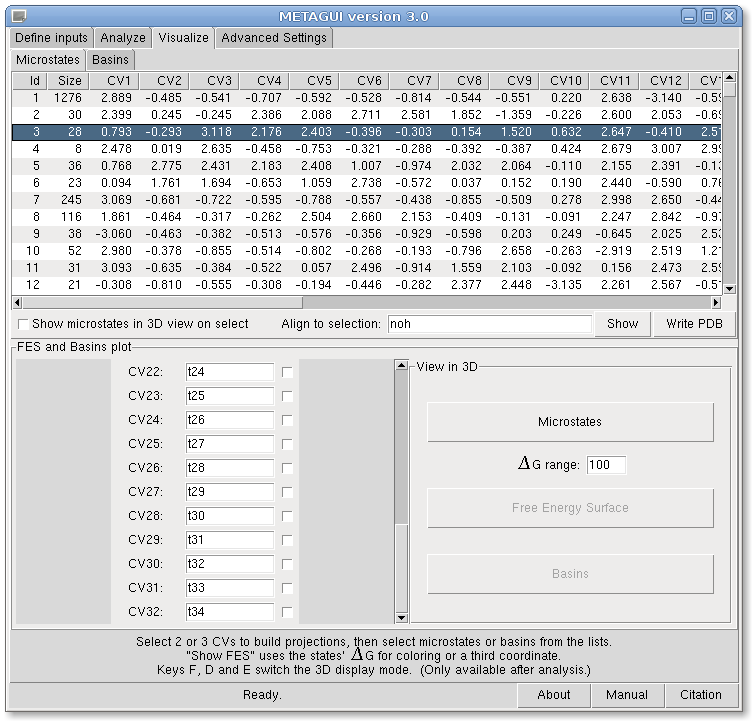}
  \caption{List of microstates obtained from the partitioning step, each
    listed with the corresponding population (\emph{Size}),
    coordinates of the centroid (\emph{CVs}), free energy
    ($\Delta G$), and cluster assignment (available after the
    Density Peak  clustering step).  }
\label{fig:visualize-microstates}
\end{figure}

\section{A benchmark on a 125 microseconds-long MD trajectory}

With the aim of illustrating the main features of METAGUI 3, we analyzed an 
unbiased simulation of Villin-headpiece folding~\cite{Lindorff-Larsen517}. 
The trajectory is 125 $\mu$s long, and includes approximately 12 folding events.
The collective variables employed were the content of $\alpha$-helix, 
the radius of gyration, the backbone dihedral distance to the crystal structure 
and the number of hydrophobic contacts. All of them were computed 
by means of PLUMED~\cite{tribello_plumed_2014} using the Plumed-GUI
interface~\cite{giorgino_plumed-gui:_2014}. 
The content of $\alpha$-helix was computed by means of the \texttt{ALPHARMSD} keyword~\cite{pietrucci_collective_2009}, 
using $r_0=1.0 \, \AAA, d_0=0.0 \, \AAA, n=6$, and $m=12$. The radius of gyration was 
computed by taking into account only the C$\alpha$ coordinates. For the backbone
dihedral distance, the reference structure considered was PDB code 2F4K~\cite{Kubelka2006546}; 
the dihedrals of terminal residues  were ignored. The number of hydrophobic contacts was computed as
the combination of the coordination coordinates between all the possible hydrophobic
residues side-chain pairs. Only the heavy atoms were considered and $r_0$ was set to 
3.5 \r{A}.

A total of 400 microstates were found by the $k$-medoids method, reducing 
the computational time by using the sieving option~\cite{shao_clustering_2007} with a maximum distance 
matrix dimension of 15,000. 
Then, the free energy of each microstate was computed 
applying the Boltzmann inversion to their populations. Finally, the free energy
wells were localized by Density Peak clustering~\cite{rodriguez_clustering_2014},
allowing the classification of the microstates in 3 different clusters.

\begin{figure}
  \centering
  \includegraphics[width=\textwidth]{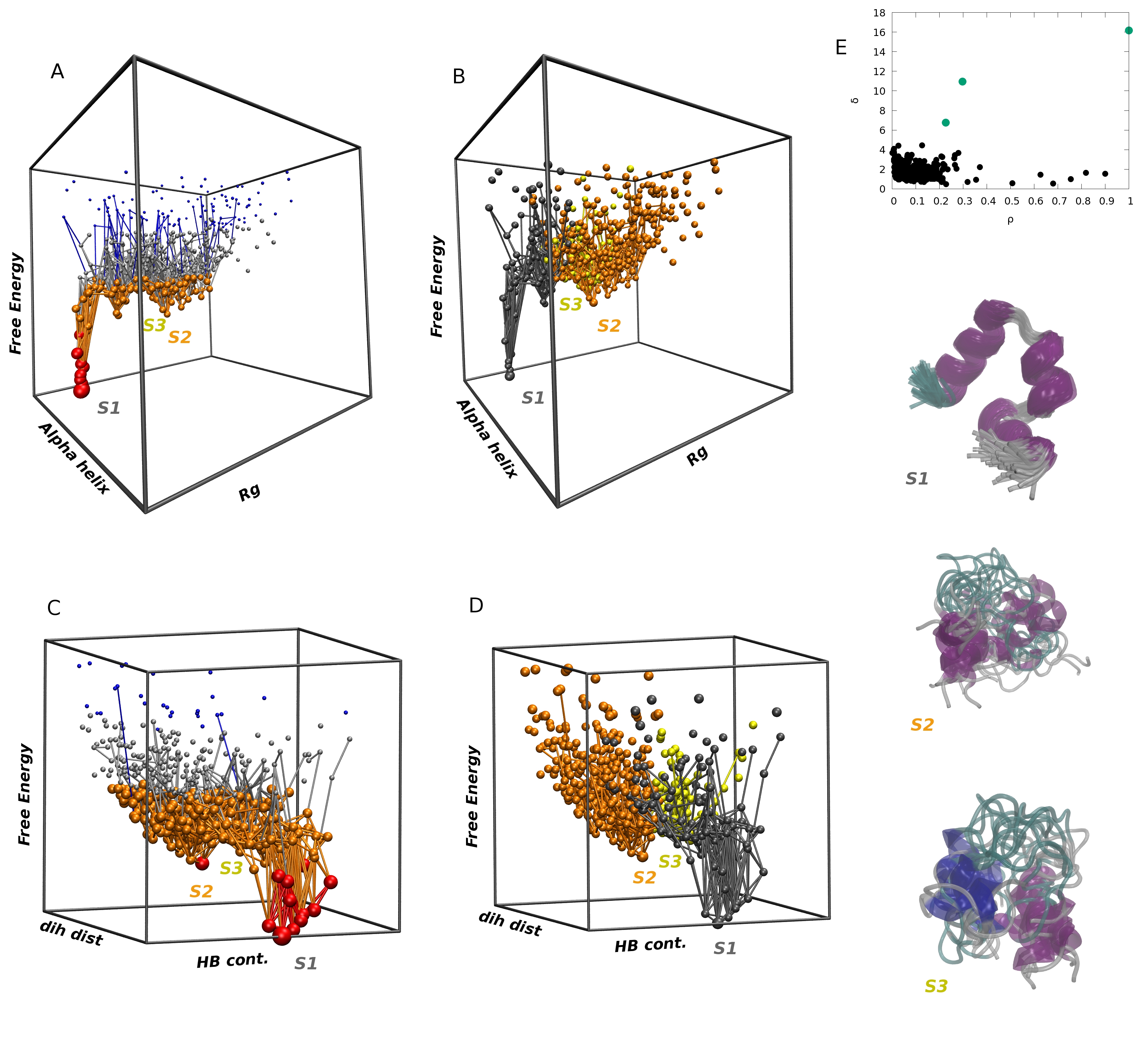}
  \caption{ METAGUI 3 output of the analysis of a 125 $\mu$s-long molecular dynamics trajectory
    of the villin headpiece:
 (A) free energy surface projected on the $\alpha$-helix and the radius of gyration CVs; 
 (B) assignment of the microstates to clusters in the projection of panel A;
 (C) free energy surface projected on the dihedral distance to the
crystal structure and the number of maintained hydrophobic contacts;
 (D) assignment of the microstates to clusters in the projection of panel C;
 (E) decision graph that leads to the cluster  partition shown in panes B and D.
 S1, S2 and S3 are the cartoon representation of the representative structures 
in each cluster.
    }
\label{fig:proj}
\end{figure}

Figure \ref{fig:proj} summarizes the results of the analysis, showing the projection on 
different CVs of the free energy surface, one using $\alpha$-helix and the 
radius of gyration (panel A) and the other using the dihedral distance to the 
crystal structure and the number of hydrophobic contacts (panel C),
the clusters partitions of the same projections (panels B and D respectively), 
the decision plane that origins the basin partition (panel E) and the 
representative structures for each basin (labeled as S1, S2 and S3). 
As expected, the microstate with minimum free energy (S1) corresponds to
the folded state. Moreover, the folding funnel is evident in both the projections. 
S2 and S3 are configurations corresponding to intermediates of the folding process. 
Remarkably, the different 
projections shown in figure  do not correspond to different calculations of
the system, but different visualizations of the same analysis. METAGUI 3  allows 
toggling between them by changing the checked CVs in the 
\emph{FES and Basins plot} frame (Figure~\ref{fig:visualize-microstates}). 
Moreover, by using the \emph{Pick} function, one can 
visualize the structures corresponding to a  microstate by simply clicking on 
a sphere in the free energy representation. 

\section{Conclusions}

The introduction of graphical tools in the MD analysis workflow can
speed-up iterations in the search for appropriate collective variables,
a necessary step towards the data-driven rationalization
of MD results.  METAGUI~3 aims at providing a quick way to
cross-reference structural and thermodynamic information, by enabling
both the selection of a set of relevant CVs on existing simulations,
and providing a quick way to define new ones. 

The source code of
METAGUI 3 is  available at the URL
\url{https://github.com/metagui/metagui3}; it can be freely
redistributed and/or modified under the terms of the GNU General
Public License (Free Software Foundation) version 3 or later.

\section{Acknowledgements}

The code of METAGUI 3.0 includes parts from the earlier METAGUI~2.0
version, written by Xevi Biarnés, Fabio Pietrucci, Fabrizio Marinelli
and Alessandro Laio.  Alessandro Laio and Alex Rodriguez 
acknowledge financial support from the grant Associazione Italiana per 
la Ricerca sul Cancro 5 per mille, Rif.\ 12214.

\bibliographystyle{model1-num-names}

\end{document}